\documentclass{aa}  

\usepackage{graphicx}
\usepackage{txfonts}
\usepackage{natbib}
\bibpunct{(}{)}{;}{a}{}{,} % to follow the A&A style
\begin{document}

   \title{Wave-particle energy transfer directly observed in an ion cyclotron wave}

   %\subtitle{Wave-particle in an ion cyclotron wave}

   \author{D. Vech \inst{1}, M. M. Martinovi\'c\inst{2}, K.~G. Klein\inst{2}, D. M. Malaspina\inst{1}, T.~A. Bowen\inst{3},  J.~L. Verniero\inst{3}, K. Paulson\inst{4}, T. Dudok de Wit\inst{5}, J.~C. Kasper\inst{6}, J. Huang\inst{6}, M.~L. Stevens\inst{4}, A.~W. Case\inst{4}, K. Korreck\inst{4}, F.~S. Mozer\inst{3}, K.~A. Goodrich\inst{3}, S.~D. Bale\inst{3}, P.~L. Whittlesey\inst{3}, R. Livi\inst{3}, D.~E. Larson\inst{3}, M. Pulupa\inst{3}, J. Bonnell\inst{3}, P. Harvey\inst{3}, K. Goetz\inst{7}
          \and
         R. MacDowall\inst{8}
          }

   \institute{Laboratory for Atmospheric and Space Physics, University of Colorado, Boulder, CO, USA\\
              \email{daniel.vech@lasp.colorado.edu}
         \and
             Lunar and Planetary Laboratory, University of Arizona, Tucson, AZ 85719, USA\\
             \and
        Space Science Laboratory, University of California Berkeley, Berkeley, CA, USA\\
        \and
        Smithsonian Astrophysical Observatory, Cambridge, MA 02138 USA\\
        \and
        LPC2E, CNRS and University of Orléans, Orléans, France\\
        \and
        Climate and Space Sciences and Engineering, University of Michigan, Ann Arbor, MI 48109, USA\\
        \and
        School of Physics and Astronomy, University of Minnesota, Minneapolis, MN, USA\\
        \and
        NASA Goddard Space Flight Center, Greenbelt, MD, USA
             }

   \date{Received DD/MM/YYYY; accepted DD/MM/YYYY}

% \abstract{}{}{}{}{} 
% 5 {} token are mandatory
 
  \abstract
   {The first studies with Parker Solar Probe (PSP) data have made significant progress toward the understanding of the fundamental properties of ion cyclotron waves in the inner heliosphere. The survey mode particle measurements of PSP, however, did not make it possible to measure the coupling between electromagnetic fields and particles on the time scale of the wave periods. }
   {We present a novel approach to study wave-particle energy exchange with PSP.}
   {We use the Flux Angle operation mode of the Solar Probe Cup in conjunction with the electric field measurements and present a case study when the Flux Angle mode measured the direct interaction of the proton velocity distribution with an ion cyclotron wave.}
   {Our results suggest that the energy transfer from fields to particles on the timescale of a cyclotron period is equal to approximately 3-6$\%$ of the electromagnetic energy flux. This rate is consistent with the hypothesis that the ion cyclotron wave was locally generated in the solar wind.}
   {}

   \keywords{solar wind -- waves -- turbulence
               }

   \maketitle
%
%-------------------------------------------------------------------

\section{Introduction}

In weakly collisional plasmas, such as the solar wind, resonant interactions between electromagnetic fields and particle distributions can lead to the transfer of energy from the fields to particles and visa versa.
For large-scale Alfv\'enic fluctuations, this exchange is largely oscillatory, dominated by undamped transfer  back  and  forth  between  fields and particles. Near ion kinetic scales,  the wave energy is irreversibly transferred to particles through secular transfer of energy \citep{howes2017diagnosing} that leads to
collisionless damping of turbulent fluctuations, which has been observed in situ \citep{chen2019evidence}.
Determining what mechanisms mediate this secular energy transfer is fundamentally important to describing the dissipation of turbulent energy in the solar wind, solar corona, planetary magnetospheres and laboratory plasmas. 
Example wave modes that may play an important role in this secular energy transfer in the solar wind include whistler, kinetic Alfv\'en (KAW) and ion cyclotron (ICW) waves.

ICWs are quasi-parallel (with respect to the local magnetic field $\mathbf{B_0}$), left-hand polarized, and have a frequency near the proton gyrofrequency. 
They can be driven by departures from local thermodynamic equilibrium, for instance large proton temperature anisotropies, $T_\perp/T_\parallel>1$ with orientations defined with respect to $\textbf{B}_0$ or field-aligned differential flows between core protons, secondary beam populations and $\alpha$-particles \citep{gary1993ion, gary2003consequences, bale2009magnetic, bourouaine2013limits, verscharen2013instabilities, verscharen2013parallel, wicks2016proton, klein2018majority, woodham2019parallel, verscharen2019multi}.

Initial observations from NASA's Parker Solar Probe (PSP \cite{fox2016solar}), the first mission to measure the solar wind below 0.3 au, revealed that ICWs are abundant in the inner heliosphere and are observed in 30-50\% of intervals with radially-aligned magnetic fields \citep{bale2019highly, bowen2020ion,bowen2020electromagnetic, verniero2020parker}. 
The lack of strong scaling of wave amplitudes with radial distance suggests that the observed ICWs are locally driven by temperature anisotropies or relative drifts. 
The survey data from the Solar Probe Analyzer for Ions (SPAN-I, 0.14 Hz cadence) and Solar Probe Cup (SPC, 1.15 Hz cadence, \cite{case2020solar}) made it possible to measure the statistical properties of proton velocity distributions during the ICW events. 
However, the survey data are not sufficiently fast to measure the coupling between electromagnetic fields and proton distributions on time scale of the wave periods, which is necessary to quantify secular energy exchange between fields and particles.

The Flux Angle operation mode (FAM) of SPC measures phase space density fluctuation in a relatively narrow range ($\approx$15 km/s wide window) of the proton velocity distributions for short intervals ($\approx$1 minute intervals 4 times a day) with up to 293 Hz cadence. 
The speed range scanned by the FAM is adjustable and it can be set to measure any part (e.g. core, beam, tail) of the proton velocity distribution functions. 
FAM data was first used to study cross helicity and residual energy at kinetic scales \citep{vech2020kinetic}.  
These initial results showed that the noise floor of the FAM for turbulent amplitudes measured during PSP's first perihelion is approximately 7 Hz, higher than the typical ion-scale spectral break of 0.3-2 Hz \citep{duan2020radial}. 
High cadence FAM data, in conjunction with the simultaneous measurements of the electric field \citep{bale2016fields}, may allow the measurement of secular energy transfer using a variation of the recently developed wave-particle correlation method of \cite{howes2017diagnosing} that we refer to as 'the standard wave-particle correlation' technique throughout this Paper. 
This technique was successfully applied to identify signatures of electron Landau damping in the turbulent terrestrial magnetosheath \citep{chen2019evidence}, but
has not yet been applied to PSP data.

This Paper presents new observations of wave-particle energy exchange involving an ICW, which was observed during the 3rd PSP perihelion at 0.23 AU. 
We measure the transfer of energy from electromagnetic fields to particles with a variation of the wave-particle correlation technique described above. Our results suggest that the energy transfer from fields to particles on the timescale of a cyclotron period is equal to approximately 3-6\% of the electromagnetic energy flux. This rate is consistent with the hypothesis that the ion cyclotron wave was locally generated in the solar wind.

%--------------------------------------------------------------------
\section{Summary of the ICW event}

\begin{figure}
 %   \figurenum{4}
    \centering\includegraphics[width=1\linewidth]{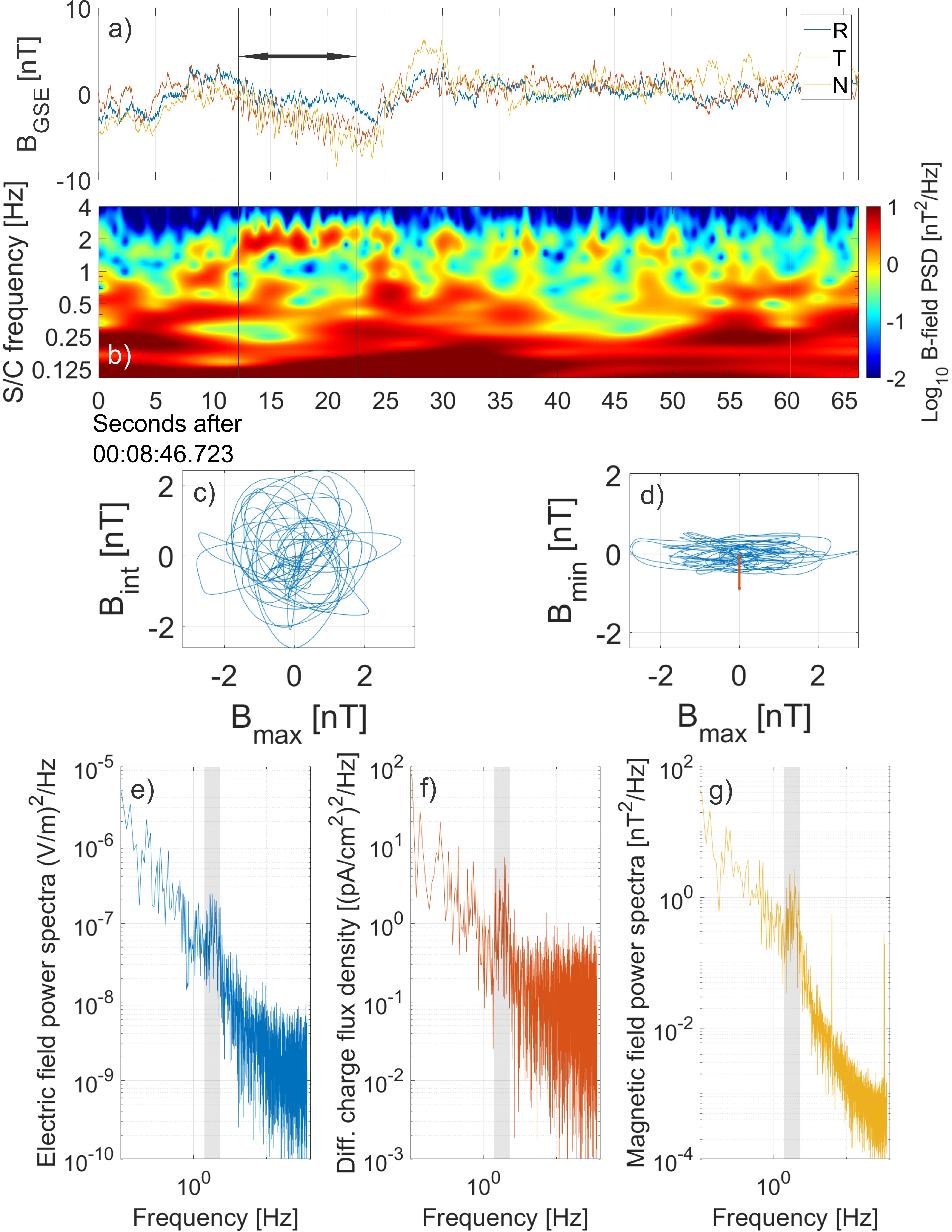}
\caption{a) Fluxgate magnetometer data (66 sec averaged subtracted) and b) trace power spectra of the magnetic field fluctuations. c-d) Hodograms showing band-pass filtered (0.6-4 Hz) fluctuations for the interval marked with an arrow in panel a) in the maximum-intermediate and maximum-minimum variance planes. The red arrow in panel d) shows $B_0$. Panels e-f) and g) show the power spectra density of electric field, differential charge flux density and magnetic field fluctuations, respectively. The frequency range with wave activity (1.4-2.3 Hz) is marked with gray shade.}
  \label{fig:wave}
\end{figure}

%-------------------------------------- Two column figure (place early!)
An ICW event was observed by FIELDS on 28-Aug-2019 00:08:58-00:09:08. 
SPC was operated in FAM with 71.4 Hz cadence between 28-Aug-2019 00:08:46-00:09:51, measuring the entire wave event. 
Fig.~\ref{fig:wave}a shows the magnetic field components for this interval in the RTN system (R  points  radially  outward  from  the  Sun,  N  is along  the  ecliptic  North  and  T  completes  the  right-hand  coordinate system) using the fluxgate magnetometer data with the 66 second average subtracted from each component. 
Fig.~\ref{fig:wave}b shows the trace power spectra of magnetic field fluctuations obtained with a continuous wavelet transform. 
Fig.~\ref{fig:wave}c and d show the results of the minimum variance analysis in the maximum-to-intermediate and maximum-to-minimum planes \citep{sonnerup1967magnetopause} for the interval marked with an arrow in Fig.~\ref{fig:wave}a. 
The minimum variance direction ( [0.93 -0.27 -0.23] in the RTN frame) is within 2$^{\circ}$ with respect to the background magnetic field ($|B_0|=$ 44.9 nT). The angle between R and the magnetic field direction ($\theta_{BR}$) was 157$^{\circ}$. The ratio of the maximum to intermediate eigenvalue is $\lambda_{max}/\lambda_{int}=1.15$ while the maximum to minimum eigenvalue is $\lambda_{max}/\lambda_{min}=17.9$, which are consistent with a circularly polarized wave. 

Fig.~\ref{fig:wave}e,f and g show the power spectra density of the electric field, differential charge flux density and magnetic field using 66 seconds of data from the FAM interval. Differential charge flux density corresponds to the current density (in units of $pA/cm^2$) on the SPC collector plates due to solar wind protons. The wave activity between 1.4-2.3 Hz is seen as enhanced wave power in all three panels. In the frequency range of the wave activity, the amplitude of the differential charge flux fluctuations is approximately one order of magnitude larger than at slightly lower (1 Hz) and higher (3 Hz) frequencies, respectively. It can be also seen that the noise floor of the FAM measurements (flattening spectra) is approximately 4 Hz. Based on these, we suggest that the FAM measurements at the frequency of the wave are not affected by noise.

We use the approach of \cite{vech2020kinetic} to calculate the RTN proton velocity fluctuations in the FAM interval and then transform them into the minimum variance frame determined previously from the magnetic field data. The phase speed of the wave is obtained with the technique of \cite{bowen2020electromagnetic} derived from Ohm's law and Faraday's law: $V_{ph}^2 = \delta V_{\perp}^2 B_{||}^2/\delta B_{\perp}^2$ where $B_{||}$ is the minimum variance component of the magnetic field, $\delta B_{\perp}$ and $\delta V_{\perp}$ are the amplitudes of the magnetic field and velocity fluctuations perpendicular with respect to $B_{||}$. We find that the average and standard deviation of $V_{ph}/V_{A}$ over the interval with the wave activity is $0.65 \pm 0.12$, which is consistent with an ion cyclotron wave (see Figure 3 in \cite{bowen2020electromagnetic}).
 
\begin{figure}
 %   \figurenum{4}
    \centering\includegraphics[width=1\linewidth]{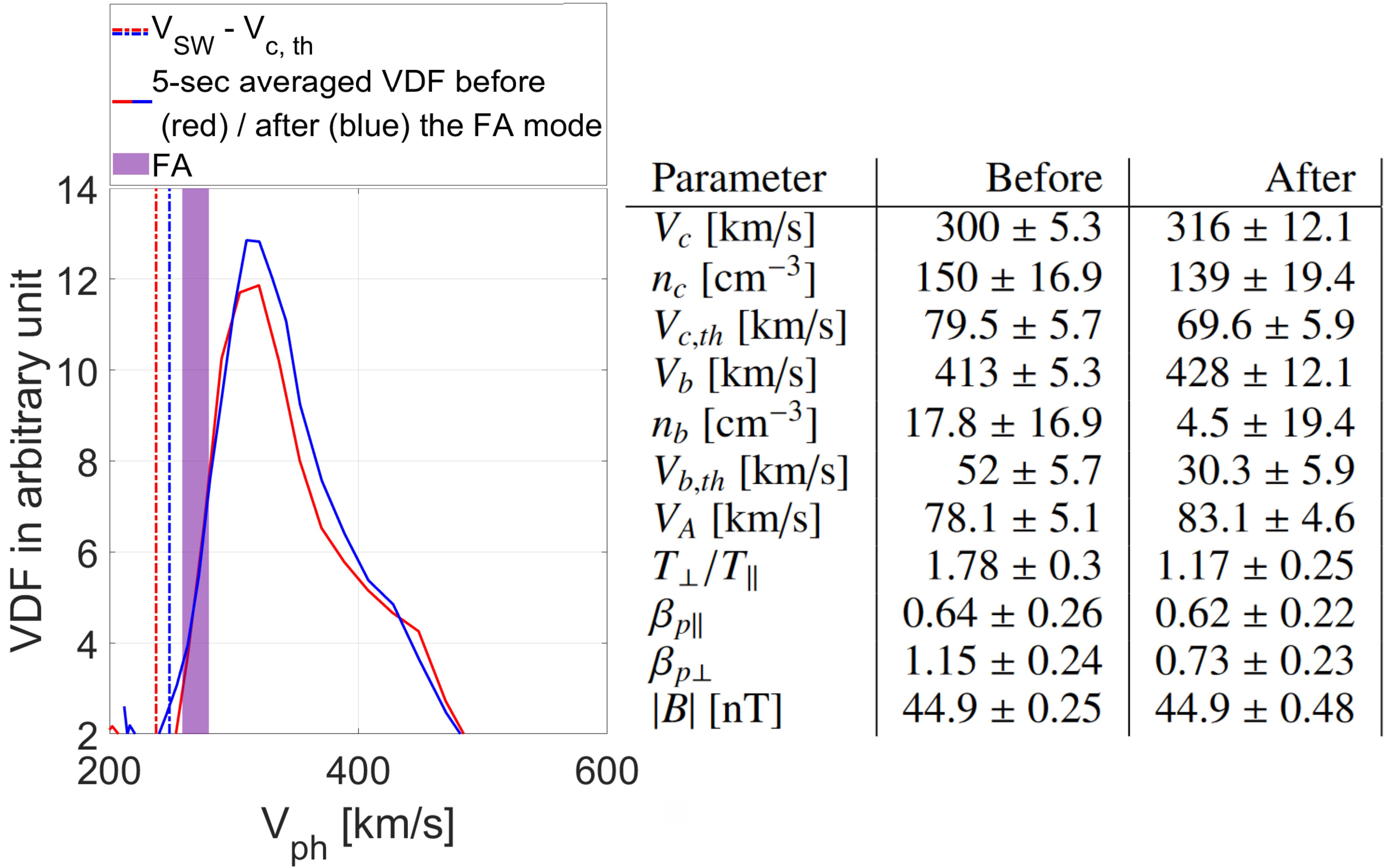}
%\plotone{figure1_landscape.eps}
\caption{5-second averages of the proton velocity distributions before (red) and after (blue) the FAM interval. The range of phase speeds measured by the FAM are shaded. The proton core thermal speed (with respect to the core proton speed) is marked with a dashed line. The table summarizes the 5-second averages and standard deviations of solar wind parameters before and after the FAM interval. In the table $c$ and $b$ subscript correspond to core and beam, respectively, and $p$ subscript correspond to fits of the entire proton distribution. The ratio of the proton thermal pressure (defined with respect to the background magnetic field) to magnetic pressure are denoted with $\beta_{||p}$ and $\beta_{\perp p}$.}
  \label{fig:FAM}
\end{figure}

Fig~\ref{fig:FAM} shows the proton velocity distributions (5 sec averages) immediately before and after the FAM interval. The shaded area shows the range of speeds (257.2 - 278.2 km/s in RTN frame) measured by the FA mode. 
As SPC measures only a single energy/charge window during a FAM, bulk solar wind parameters are not available. 
The table in Fig~\ref{fig:FAM} compares the core plasma parameters before and after the FAM interval. We find less than 15\% change in the core proton values, and that the flow angle was closely aligned (9.3$^{\circ}$) with the normal of the SPC collector plate. These results suggest that SPC measured approximately the same part of the velocity distribution in the FAM and there were no significant fluctuations in the core parameters. The proton beam density and temperature showed larger variability, changing by a factor of 3.95 and 1.71, respectively. The proton temperature anisotropy ( $T_{\perp}/T_{||}$) was 1.78 and 1.17 before and after the FAM interval, respectively, which are consistent with an Alfv\'en-cyclotron instability. Proton temperature anisotropy measurements made with SPC are determined by examining the change in the apparent flow angle across the peak of the observed velocity distribution function. This measurement is made separately in the x-z and y-z planes in spacecraft coordinates, and the final value is a combination of the two weighted by uncertainty and assuming gyrotropy. For details of this approach see \cite{paulson2020}.

%\begin{table}[h!]
%\centering
%    \label{tab:table1}
%   \begin{tabular}{l|r|r|r|r} % <-- Changed to S here.
%      Parameter & Before & After \\
%     %$\alpha$ & $\beta$ & $\gamma$ & asd\\
%      \hline
%      $V_{c}$ [km/s] & 300 $\pm$ 5.3 & 316 $\pm$ 12.1\\
%      $n_{c}$ [cm$^{-3}$] & 150 $\pm$ 16.9 & 139 $\pm$ 19.4\\
%      $V_{c, th}$ [km/s] & 79.5 $\pm$ 5.7 & 69.6 $\pm$ 5.9\\
%      $V_{b}$ [km/s] & 413 $\pm$ 5.3 & 428 $\pm$ 12.1\\
%      $n_{b}$ [cm$^{-3}$] & 17.8 $\pm$ 16.9 & 4.5 $\pm$ 19.4\\
%      $V_{b, th}$ [km/s] & 52 $\pm$ 5.7 & 30.3 $\pm$ 5.9\\
%      $V_A$ [km/s] & 78.1 $\pm$ 5.1 & 83.1 $\pm$ 4.6\\
%      $T_{\perp}/T_{||}$ & 1.78 $\pm$ 0.3 & 1.17 $\pm$ 0.25\\
%      $\beta_{p||}$ & 0.64 $\pm$ 0.26 & 0.62 $\pm$ 0.22\\
%      $\beta_{p\perp}$ & 1.15 $\pm$ 0.24 & 0.73 $\pm$ 0.23\\
%      $|B|$ [nT] & 44.9 $\pm$ 0.25 & 44.9 $\pm$ 0.48\\
%    \end{tabular}
%        \caption{5-second averages and standard deviations of solar wind parameters before %and after the FA mode interval.}
%\end{table}

\section{Quantifying Rate of Energy Transfer}

The standard wave-particle correlation technique relies on three-dimensional velocity distributions and three-dimensional measurements of the electric field. 
However, SPC measures only the reduced velocity distribution in the plane of the collector plate. 
The FIELDS instrument measures the radial electric field fluctuations only for high frequencies (above kHz range) and thus low frequency E-field is only available in the plane of the heat shield (e.g. T and N components). 
For the analysis of PSP data, and in particular for the velocity-range limited FAM observations, the wave-particle correlation technique
needs to be modified to estimate the energy transfer between fields and particles.

First, we calculate the Poynting flux of the electromagnetic field by using the T and N components of the $\mathbf{E}$ and $\mathbf{B}$ fields
\begin{equation}
P = (0,E_T,E_N) \times (0,B_T,B_N)/\mu_0 [W/m^2]
\end{equation}
and the energy flux of the solar wind protons in the range of the FAM is 
\begin{equation}
K = \frac{1}{2}\cdot m_p \cdot v_0^2 \cdot \phi_p [W/m^2],
\end{equation}
where {$\phi_p$ is the proton flux} [$\#/(m^2 \cdot s)$] measured by the FAM, v$_0 = -38.8$ km/s is the center of the FAM speed range with respect to the average core proton speed (see Fig~\ref{fig:FAM}) and $m_p$ is the mass of a proton.

\begin{figure}
 %   \figurenum{4}
    \centering\includegraphics[width=1\linewidth]{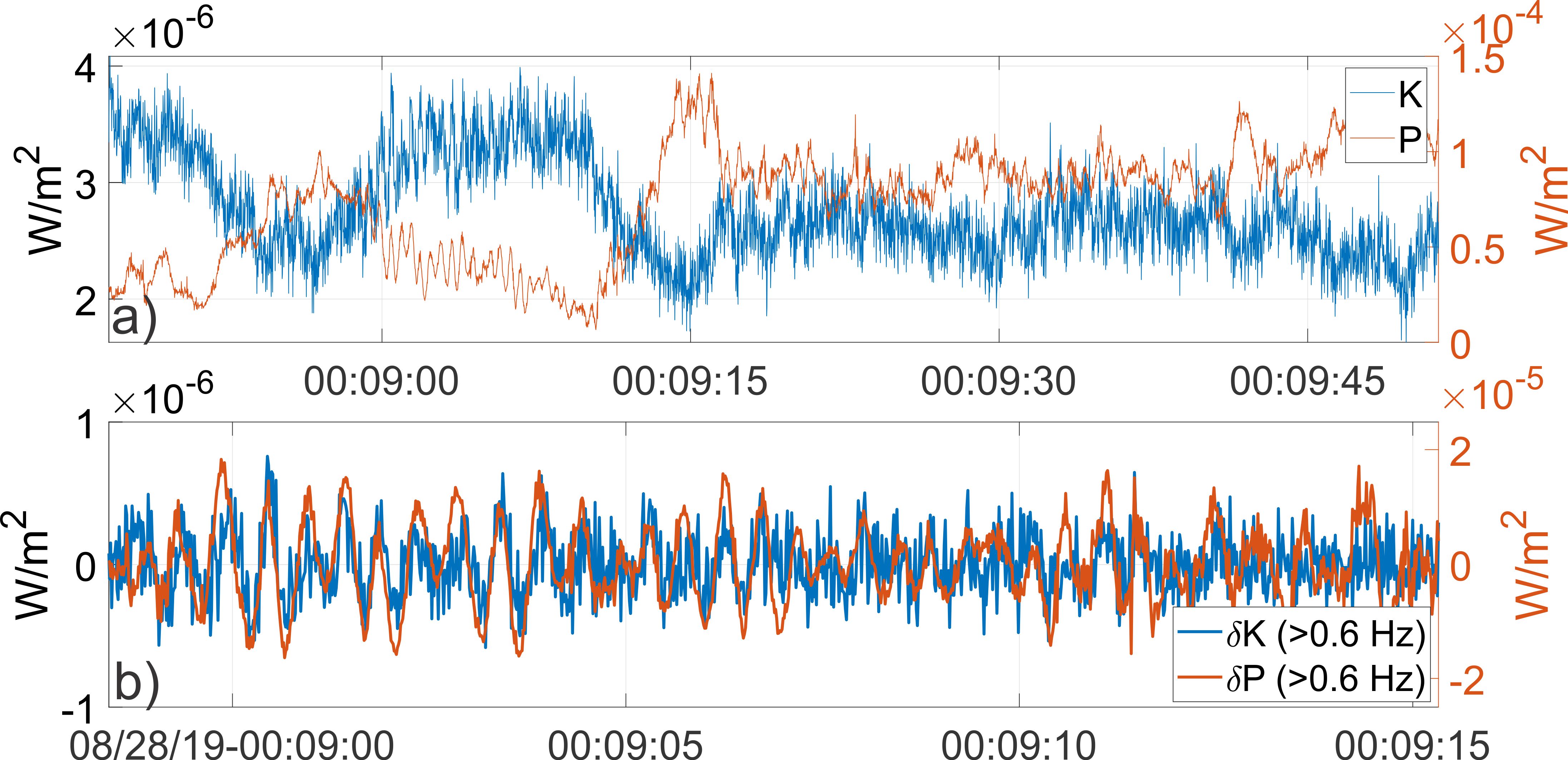}
%\plotone{figure1_landscape.eps}
\caption{a) P and K for the entire studied interval, the magnitude of P is larger by a factor of 20-50 than K and b) the high frequency ($>$0.6 Hz) fluctuations for the sub-interval with the wave activity.}
  \label{fig:PK}
\end{figure}

Fig~\ref{fig:PK}a shows the derived K and P parameters for the entire selected interval while Fig~\ref{fig:PK}b focuses on the sub-interval with the ICW and shows the high pass filtered (at 0.6 Hz) P and K parameters, $\delta P$ and $\delta K$. The remarkably strong correlation between $\delta K$ and $\delta P$ suggests that significant transfer of energy between the fields and particles may take place in the ICW. We note that measuring high correlation between fields and particles is necessary but not sufficient condition for wave damping (see \cite{gershman2017wave}) and the secular energy transfer will be tested with the following analysis.

To quantify the transfer of energy between the electric field and solar wind protons within the FAM energy range, we modify the approach of \cite{howes2017diagnosing} to the following expression 

\begin{equation}
C_{\textrm{FAM};\parallel,\perp} = q_p v_{0} \delta E_{\parallel, \perp}
\delta \phi_p
\end{equation}
with the time-integrated value
$\int C_{\textrm{FAM}}(t) dt$ representing the transfer of energy flux, in $[W/m^2]$, mediated by protons with energies within the FAM range. In Equation 3 $q_p$ is the charge of a proton, $E_{||}$ and  $E_{\perp}$ correspond to the electric field fluctuations parallel and perpendicular with respect to the magnetic field in the T-N plane. The direction of $E_{\perp}$ is defined by the cross product of the +R unit vector and the magnetic field in the T-N plane (0, B$_T$, B$_N$).

The proposed technique allows us to identify whether the energy exchange is preferentially coupled to $E_{||}$ or $E_{\perp}$. We note that in order to identify the secular energy transfer both $\phi_p$ and $E_{||, \perp}$ are high pass filtered at 0.6 Hz to remove any constant phase space density and electric field structures (hence the $\delta$ notation in Equation 3). Similar filtering technique was used by \cite{chen2019evidence} as well. For our analysis the electric field data measured in the spacecraft frame (${E}_{sc}$) was converted into the frame of the solar wind: $\mathbf{E}_{sw} = \mathbf{E}_{sc} + \mathbf{v}_{sw} \times \mathbf{B} $ \citep{chen2011frame}. For the calculation of $\mathbf{v}_{sw}$ during the FAM interval, we used the approach of \cite{vech2020kinetic}.

We calculate $\int C_{\textrm{FAM}}(t) dt$ for 3-second blocks of the data. The length of this interval ensures that multiple ($\approx$6) wave periods are included. In the absence of secular energy transfer $\int C_{\textrm{FAM}}(t) dt$ is zero over multiple wave periods since the oscillatory transfer of energy between fields and particles averages out each other. We adjust the measured $\int C_{FAM ||; \perp} dt$ values by multiplying with the ratio of the cyclotron period ($2\pi/\Omega_p=1.46$ sec) and the length of the integration window (3-sec) therefore the computed values correspond to the average transfer of energy flux on the time scale of a cyclotron period.

We test the statistical significance of the measured $\int C_{\textrm{FAM}}(t) dt$ with the phase-randomized technique suggested by \cite{howes2017diagnosing}: the electric field data is Fourier transformed, a uniformly distributed random phase is added and the data is inverse Fourier transformed. We calculate $\int C_{\textrm{FAM}}(t) dt$ 40 times for each 3-second block of data with the randomized electric field data and obtain a distribution of the possible values.

\begin{figure}
 %   \figurenum{4}
    \centering\includegraphics[width=1\linewidth]{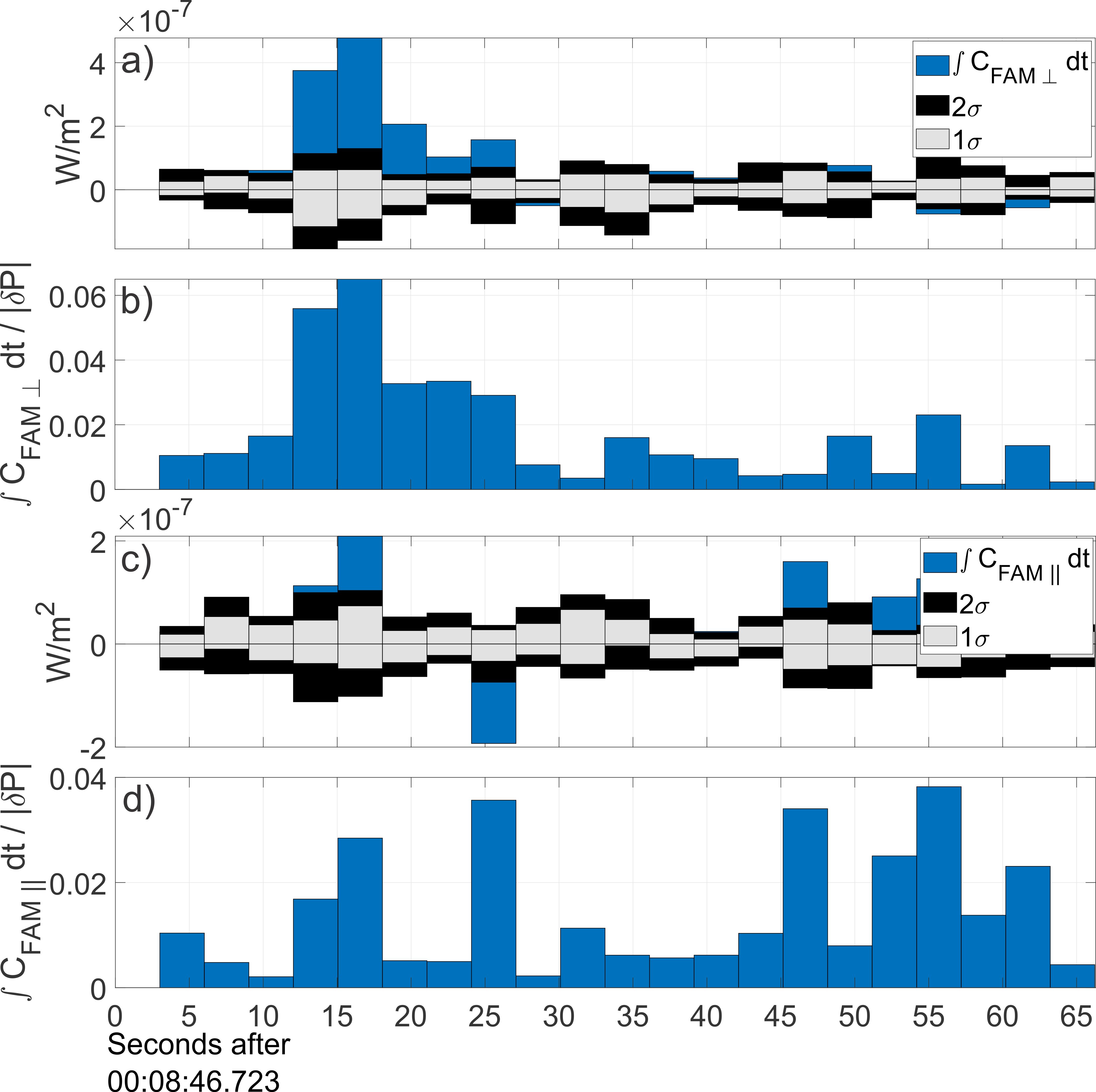}
%\plotone{figure1_landscape.eps}
\caption{a)-c) $\int C_{FAM ||; \perp} dt$. The gray and black bars show the 1 and 2$\sigma$ values of $\int C dt$ calculated with the synthetic (randomized) electric field data. b)-d) Ratio $\int C_{FAM ||; \perp} dt$ and the average of $| \delta P|$ fluctuations in each 3-second blocks of data.}
  \label{fig:4}
\end{figure}

The results of the analysis are presented in Fig~\ref{fig:4}. In panels a) and c) the gray and black bars correspond to the 1 and 2$\sigma$ distributions of $\int C_{\textrm{FAM}}(t) dt$ calculated with randomized electric field data, respectively and the blue bars show the $\int C_{FAM ||; \perp} dt$ values calculated with the real electric field data. It can be seen that there is a 12-second interval where the $\int C_{FAM \perp} dt$ values are statistically significant on the 2$\sigma$ level, which overlaps with the wave activity presented in Fig~\ref{fig:wave}a. The positive sign of $\int C_{FAM ||; \perp} dt$ is consistent with transfer of energy flux from electromagnetic fields to particles \citep{ chen2019evidence, klein2020diagnosing}. For the same interval the $\int C_{FAM ||} dt$ values were statistically significant for only two 3-second segments of data. At around the 25th second both $\int C_{FAM ||} dt$ and $\int C_{FAM \perp} dt$ are statistically significant, however it is likely caused by a magnetic discontinuity (see change of the magnetic field components in Fig~\ref{fig:wave}a) and not by an ICW.

In order to understand the significance of the measured $\int C_{FAM ||; \perp} dt$ values during the wave event, we compare them to the average magnitude of the high pass filtered (at 0.6 Hz) P parameter ($|\delta P|$) in each 3-second bin. The ratio of $\int C_{FAM \perp} dt$ and $|\delta P|$ suggests that the exchange of energy flux between $E_{\perp}$ and particles on the time scale of a cyclotron period is approximately 3-6\% of the amplitude of Poynting flux fluctuations while it is approximately 3\% for $E_{||}$. These results correspond to $\gamma/\Omega_p \in$ [0.032; 0.066]. Due to the limitation of the measurements (2-D plane geometry), the derived damping rates should be considered as lower thresholds. However, we suggest that the bulk of the energy transfer is still captured since the largest amplitude electric field fluctuations are approximately in the T-N place and the largest amplitude particle flux fluctuations are in the R component. 

The full equation of Poynting theorem includes three terms: $\textbf{j} \cdot \textbf{E}$ (represented by $\int C_{\textrm{FAM}}(t) dt$), Poynting flux term (represented by P) and $\partial/\partial t(\epsilon|E|^2+|B|^2/\mu)$. We suggest that the latter term has negligible amplitude compared to the Poynting term due to the fact that the ICW is a highly incompressible structure where fluctuations in the magnitude of the electromagnetic field are significantly smaller than the components perpendicular to the wave propagation. We tested this assumption for the magnetic field (where all three components are available) and found that $dB_{\perp}/\delta|B| \approx$ 50-200 at the wave frequency. The $\partial/\partial t(\epsilon|E|^2+|B|^2/\mu)$ term also includes secular changes in the amplitude of E and B on quasilinear timescales related to $\gamma/\Omega_p$. We found that the power spectra based on the T and N components of E and B fields has factor of $\approx$ 100 larger amplitude than the power spectra of |E| and |B| (both measured in the T-N plane). Therefore we quantify the field-particle energy transfer with Equation 3 and the Poynting term.

\section{Linear stability analysis of the ICW event}

Using the \texttt{PLUME} numerical solver \citep{klein2015plume} 
we calculate linear dispersion relations for the plasma parameters before the FAM. \texttt{PLUME} calculates the linear normal mode response for an arbitrary number of relatively drifting, bi-Maxwellian ion and electron populations. We model the plasma with separate proton core and beam components and a single electron distribution, using the measurements from the "before" interval (see Figure 2) to yield dimensionless parameters $\beta_{\parallel,c}=0.947$,
$V_{\parallel,c,th}=2.265 \times 10^{-4}c$, 
$T_{\parallel,b}=0.43 T_{\parallel,c}$,
$n_{b}=0.11 n_{c}$, and
$T_{\parallel,e}=1.00T_{\parallel,c}$. The value of $\beta_{\parallel,c}$ is based on the ratio of the thermal pressure associated with the core proton distribution (where the core and beam are Maxwellian fits to the SPC measurements of the proton velocity distribution) to the magnetic pressure, rather than the total proton thermal pressure extracted from an analysis of the moments ($\beta_{p\parallel}$ parameter listed in Figure 2). The relative drift speed between the beam and core was set to $1.41 V_{A}$.

Using the method described in \cite{paulson2020}, the total proton temperature anisotropy was found to be $T_{\perp,p}/T_{\parallel,p} =1.79$. Given the uncertainties in this method, and in disentangling anisotropies in the core from the total anisotropy, we consider solutions with $T_{\perp,c}/T_{\parallel,c}=1.0, 1.4, $ and $1.8$. 
The electrons and proton beam were assumed to have isotropic temperatures.

\begin{figure}
    \centering
    \includegraphics[width=\columnwidth]{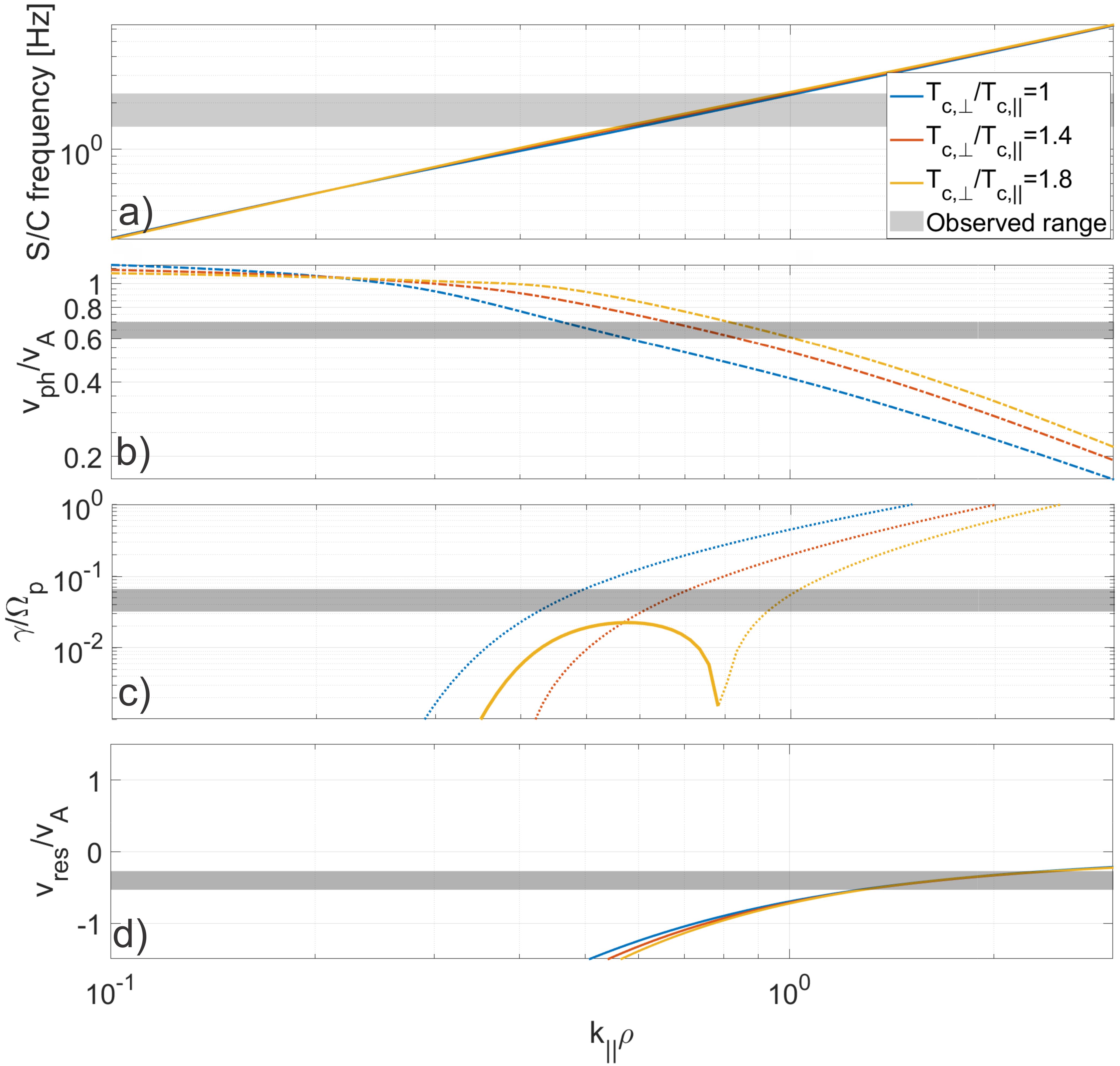}
    \caption{Overview of the \texttt{PLUME} dispersion solution for forward propagating Alfv\'en/IC modes for varying $T_{\perp,c} / T_{\parallel,c}$.
    Panel a) shows the Doppler shifted frequency, b) shows the phase speed normalized by the Alfv\'en speed. Panel c) shows the normalized growth (solid line) or damping (dashed) rate $\gamma/\Omega_p$, which strongly depends on the temperature anisotropy. Panel d) shows the ratio of the resonant velocity normalized by the Alfv\'en speed.
    }
    \label{fig:PLUME}
\end{figure}

We identify both the forward and backward propagating Alfv\'en waves at $k_\perp \rho_c= k_\parallel \rho_c=10^{-3}$ (where $\rho_c$ is the proton gyroradii; $k_\perp$ and $k_\parallel$ are the parallel and perpendicular wavenumbers defined with respect to $\mathbf{B_0}$), and trace these normal mode solutions to larger $k_\parallel$ for fixed $k_\perp$. The backwards propagating solutions, with cyclotron resonant velocities greater than the core proton velocity are found to be stable for all three anisotropies considered. The characteristics of the forward solutions are shown in Fig~\ref{fig:PLUME} where we make direct comparison with the wave parameters (shaded areas in each panel) derived in the previous sections.

Fig~\ref{fig:PLUME}a shows the Doppler shifted wave frequency as a function of $k_{||}\rho_c$. The gray shade corresponds to the observed frequency of the wave (see Figure 2e-f-g), which constrains the range of $k_{||}\rho_c$ values to approximately $k_{||}\rho_c \in$[0.6; 1]. In Fig~\ref{fig:PLUME}b and c, the computed $V_{ph}/V_A$ ratio (see Section 2) and the empirically derived $\gamma/\Omega_p$ ratio (see Section 3) are consistent with the same range of $k_{||}\rho_c$ values. In Fig~\ref{fig:PLUME}c the intersection of the shaded area with the line plots suggests that the wave undergoes damping for all three anisotropy values considered here.

In Fig~\ref{fig:PLUME}d, we calculate the resonant cyclotron velocities $v_{\textrm{res}}$ as
\begin{equation}
    v_{\textrm{res}} = \frac{\omega_{\textrm{r}} -n\Omega_p}
    {k_\parallel}
\end{equation}
where $\omega_r$ is the real frequency, $n=+1$ for forward propagating Alfv\'enic solutions, as appropriate for their left-handed plasma frame polarizations. As the cyclotron resonant velocities for backward-propagating ICWs, with $n=-1$ due to their right-handed plasma-frame polarization (not shown), are at velocities greater than $V_{c}$, they are not resolved by this particular FAM observation. The velocity space observed by the FAM covers $v \in [- 0.53; -0.27]V_{A}$ below $V_{c}$. Although the FAM range does not overlap with the $v_{\textrm{res}}/V_A$ curves in the range of $k_{||}\rho_c=$[0.6; 1], the gap between them is only 13 km/s (0.16$V_A$) at $k_{||}\rho_c=$1.

Given the sign of the energy transfer measured by $\int C_{FAM \perp} dt$ and the comparison of the \texttt{PLUME} results with observations in Fig~\ref{fig:PLUME}, we suggest that our observations are consistent with an ICW being generated by a nearby region of plasma with a larger temperature anisotropy, which can then propagate to a region of lower anisotropy, where it can be efficiently absorbed. It is likely this last step that the FAM is observing.

\section{Conclusions}

In this Paper we described a method to measure the direction and magnitude of the energy flux transfer between electromagnetic fields and particles in the range of the Flux Angle mode of SPC. Measurements of the Doppler shifted wave frequency, phase speed and the $\gamma/\Omega_p$ damping ratio were in good agreement with the properties of an ICW with $k_{||}\rho_c=$[0.6; 1]. Our results are consistent with damping of an ICW where the energy flux is transferred from electromagnetic fields to the particles with a magnitude of approximately 3-6\% (per cyclotron period) of the Poynting flux fluctuations.

The results make it possible to clarify where (local vs. solar origin) the ICW was generated: at the time of the observation $\delta B/B_0 \approx 0.1$. If we assume that the decay rate of $\delta B$ is in the range of 3 and 6\% per cyclotron period and that $\delta B/B_0 <1$ in an Alfv\'en wave, then the upper thresholds of the wave generation are 34.8 (50-sec) and 71.6 (104-sec) cyclotron periods before PSP observed the waves, respectively. These timescales are consistent with the hypothesis that the ICW was locally generated in the solar wind.

\begin{acknowledgements}
KGK was supported by NASA Grant 80NSSC19K0912. The authors thank the Parker Solar Probe team, especially the FIELDS and SWEAP teams for their support. The FIELDS experiment on the Parker Solar Probe spacecraft was designed and developed under NASA contract NNN06AA01C. TD acknowledges support from CNES. All data used in this work are available on the FIELDS data archive: http://fields.ssl.berkeley.edu/data/ and the SWEAP data archive: http://sweap.cfa.harvard.edu/pub/data/sci/sweap
\end{acknowledgements}

% for the bibliography, at the end
%\bibliographystyle{aa} % style aa.bst
%\bibliography{cs} % your references Yourfile.bib


\end{document}